# Monitoring single membrane protein dynamics in a liposome manipulated in solution by the ABELtrap

T. Rendler[a], M. Renz[a], E. Hammann[a], S. Ernst[a], N. Zarrabi[a], M. Börsch*[a]

[a] 3rd Institute of Physics, University of Stuttgart, Pfaffenwaldring 57, 70550 Stuttgart, Germany

## ABSTRACT

$F_oF_1$-ATP synthase is the essential membrane enzyme maintaining the cellular level of adenosine triphosphate (ATP) and comprises two rotary motors. We measure subunit rotation in $F_oF_1$-ATP synthase by intramolecular Förster resonance energy transfer (FRET) between two fluorophores at the rotor and at the stator of the enzyme. Confocal FRET measurements of freely diffusing single enzymes in lipid vesicles are limited to hundreds of milliseconds by the transit times through the laser focus. We evaluate two different methods to trap the enzyme inside the confocal volume in order to extend the observation times. Monte Carlo simulations show that optical tweezers with low laser power are not suitable for lipid vesicles with a diameter of 130 nm. A. E. Cohen (Harvard) and W. E. Moerner (Stanford) have recently developed an Anti-Brownian electrokinetic trap (ABELtrap) which is capable to apparently immobilize single molecules, proteins, viruses or vesicles in solution. Trapping of fluorescent particles is achieved by applying a real time, position-dependent feedback to four electrodes in a microfluidic device. The standard deviation from a given target position in the ABELtrap is smaller than 200 nm. We develop a combination of the ABELtrap with confocal FRET measurements to monitor single membrane enzyme dynamics by FRET for more than 10 seconds in solution.

**Keywords:** Rotary motor, $F_oF_1$-ATP synthase, Anti-Brownian electrokinetic trap ABELtrap, single-molecule detection, Förster resonance energy transfer FRET

## 1 INTRODUCTION

Functionality of living cells is based on the consumption of ATP (adenosine triphosphate). To produce this chemical energy currency molecule ATP, a complex biological nanomachine $F_oF_1$-ATP synthase is working in all types of cells. This membrane-embedded enzyme consists of two distinct parts which are operating as two coupled rotary motors[1-3]. In the plasma membranes of *Escherichia coli* bacteria, the proton-driven $F_o$ motor is powering the ATP-synthesizing $F_1$ part. To analyze the conformational dynamics of this enzyme, we apply Förster resonance energy transfer, FRET (or **f**luorescence **r**esonance **e**nergy **t**ransfer, respectively) between two[4-16] or three[17] specific amino acid positions marked with single fluorophores. Following membrane protein purification procedures and fluorescence labeling, the membrane enzyme has to be reconstituted into a lipid bilayer (liposome) to become functional for ATP synthesis. These lipid vesicles contain only a single enzyme. They are called proteoliposomes and have been carefully characterized for their biochemical activities. Reviews exist which briefly describe the experimental details for FRET-labeling $F_oF_1$-ATP synthase and how to measure the catalytic turnover rates[18-21].

Based on the FRET methodologies to study $F_oF_1$-ATP synthase we also investigate the conformational dynamics of other ATP-driven membrane transporters on the single-molecule level, that is, the $K^+$ transporter KdpFABC[22, 23], the vacuolar proton pump $V_oV_1$-ATPase[24-27], the ABC transporter P-glycoprotein [S. Ernst, B. Verhalen et al, Proceedings of SPIE, to be published in 2011] and the bacterial protein-translocating membrane machinery SecA-YEG. All transporters are specifically labeled at protein domains which move during function, and they are reconstituted into liposomes with diameters of about 100 to 200 nm.

.................................................................................................................................................................

* m.boersch@physik.uni-stuttgart.de; phone (49) 711 6856 4632; fax (49) 711 6856 5281; http://www.m-boersch.org

Single-molecule FRET measurements take place in a buffer droplet which is placed on a confocal microscope of custom, variable design (including multiple lasers, piezo-driven sample scanning and a spectrograph)[25, 28-43]. When a proteoliposome with a FRET-labeled membrane protein eventually passes the confocal detection volume due to Brownian motion, conformational changes of the active enzyme are observed as fluctuations of the FRET efficiencies during the transit times. These transit times limit the experimental access to the internal dynamics of the membrane proteins, and thus we are interested in increasing the observation times. We compared different trapping techniques that are suitable for our artificial biological systems. Here we report on the conditions to use an optical trap for holding a proteoliposome, and we describe the achieved progress for a combination of the Anti-Brownian electrokinetic trap (***ABELtrap***, invented by A. E. Cohen and W. E. Moerner[44-48]) with the confocal FRET measurements of single membrane protein dynamics.

## 2 BROWNIAN MOTION AND TRAP FORCES

### 2.1 Brownian motion

The diffusion of a mesoscopic object in solution is a consequence of impacts between thermally excited solution molecules and the suspended object itself. This phenomenon is called Brownian motion. The characteristic quantity to describe the diffusion of a spherical shaped vesicle is the diffusion coefficient $D$ which depends on the equilibrium temperature $T$ of the suspension, the Boltzmann constant $k_B$ and a damping constant $\gamma$ of the surrounding medium:

$$D = \frac{k_B T}{\gamma} \qquad (1)$$

To describe the movement of a diffusing particle with mass $m$ in one direction $x$ under the influence of a force $F$ we use the Langevin-equation:

$$m\ddot{x} = -\gamma \dot{x} + \Gamma_{st}(t) + F \qquad (2)$$

with $\Gamma_{st}(t)$, the stochastic force causing diffusion. Using the definition $\tau = m/\gamma$ we get in the limit of over-damping (i.e $t/\tau \gg 1$) and constant forces a simple solution for eq. (2):

$$x(t) = \frac{1}{\gamma} F t + x_s \qquad (3)$$

with $x_s$, the displacement caused by the stochastic force. Using the Fokker-Planck-Equation for a freely diffusing particle we write the probability density function $p$ for the displacement $x_s$ after time $t$:

$$p(x_s, t) = \frac{1}{\sqrt{2\pi\sigma^2}} \cdot exp\left(-\frac{x_s^2}{2\sigma^2}\right) \qquad (4)$$

The variance $\sigma^2$ then equals $2Dt$. Equations (3) and (4) were used to simulate the particle in optical tweezers (see below).

### 2.2 Forces in Optical Tweezers

Optical tweezers are based on the principle that light and matter of particles can interact and form an attractive potential depending on special conditions. This force $F_{opt}$ can be described as:

$$F_{opt} = \frac{Q \cdot n_m \cdot P}{c} \qquad (5)$$

with $n_m$, refractive index of the surrounding medium, $P$, laser power of the trapping laser, $c$, speed of light, and $Q$, a quantity characterizing the efficiency of the trap [49]. Assuming that our particles are small relative to the wavelength of the trapping laser light (that is, the dipole approximation or the Rayleigh regime, respectively), the forces acting on the particle can be divided into two parts: the scattering force $F_{scatt}$ and the gradient force $F_{grad}$ [50, 51]. The scattering force originates from the absorption of the laser light impulse with a cross section σ, and this is a repulsive force. Its average impulse points along the propagation axis of the light beam $\vec{k}_0$ with intensity $I$:

$$\vec{F}_{scatt} = n_m \frac{\vec{k}_o \cdot I \cdot \sigma}{c} \tag{6}$$

The gradient force can be attractive or repulsive depending on the refractive index of the particle with respect to the surrounding medium or solution, respectively. It depends on the polarizability α of the particle and the intensity gradient $\vec{\nabla} I$ of the laser beam:

$$\vec{F}_{grad} = \frac{2\pi\alpha}{c \cdot n_m^2} \cdot \vec{\nabla} I \tag{7}$$

If the refractive index of the solution is smaller than the refractive index of the particle the gradient force creates an attractive potential. We will work only in this regime to trap proteoliposomes (~ 130 nm diameter) in aqueous buffer.

**2.3 Forces in the ABELtrap**.

The ABELtrap is an electrokinetic trap for particles down to nanometer sizes[48]. There movement is limited to two dimensions in microfluidic devices with a height of less than 1 micrometer[44, 52]. Electric fields are generated by electrodes which push the charged particle towards a predefined target position. Depending on the distance from the target position, different voltages are applied. Here we use fluorescently labeled liposomes which have to be pushed to a fixed confocal laser excitation volume to measure FRET, and, accordingly, to observe the functional dynamics of the liposome-embedded membrane protein.

There are several ways to identify and localize a fluorescent proteoliposome. We choose a fast EMCCD camera with single photon detection sensitivity. To correct the position of a proteoliposome, electrical fields are generated by platinum electrodes in the ABELtrap. The interactions between the electrical field and the particles are called electrokinetic effects and can be generalized to the relationship:

$$\Delta x_p = k \cdot E \cdot \Delta t \tag{8}$$

Here $\Delta p_x$ is the distance the particle has moved during the time interval $\Delta t$ during the application of an electric field $E$ along the pathway. The factor $k$ is assumed to be constant during the trap measurement. It is not important to know the exact value of $k$ before beginning the ABELtrap experiment, because the appropriate value is found by maximizing the trapping efficiency of the ABELtrap empirically. The trapping efficiency of the ABELtrap depends on the diffusion constant $D$ of the particle, the localization precision and the shape or geometry of the trapping chamber.

For a freely diffusing particle the diffusion length $d_\tau$ for one dimension within the time interval $t$ is given by:

$$d_\tau = \sqrt{2Dt} \tag{9}$$

Moving the particle back after time $t$ to its starting position and repeating this for several time intervals, we find that the particle position is Gaussian distributed. The distribution gets broadened because we cannot perfectly localize the particle due to noise, a limited pixel size and "blooming". Thus the measured position distribution will always be broader then the real position distribution of the particle. In addition, the chamber design of the ABELtrap can cause a broadening of the distribution. All broadening effects are combined in a variance $\sigma_\Delta$. The overall variance $\sigma_m$ of the measured position distribution results:

$$\sigma_m^2 = \sigma_\Delta^2 + d_\tau^2 \tag{10}$$

# 3 TRAPPING PROTEOLIPOSOMES WITH OPTICAL TWEEZERS

## 3.1 Setup of the microscope

Given the small size of the proteoliposomes and a very low change in the refractive indices between liposome and buffer, we expected that optical tweezers operating at low laser intensities are not suitable for trapping a small and sensitive proteoliposome. To verify this, we combined a confocal microscope for single-molecule FRET measurements with optical tweezers (Figure 1). As the test sample "green fluorescent" polystyrene beads with a diameter of 100 nm (Invitrogen) were illuminated by a pulsed laser at 488 nm with a few nW (PicoTA 490; Picoquant, Berlin), and bead fluorescence was detected in the spectral range between 498 nm to 568 nm (interference filter BP 532/70, AHF Tübingen) using a single photon counting avalanche photodiode ("APD2" in Figure 1; SPCM-AQR-14, Perkin Elmer). Fluorescence time trajectories of the beads were recorded in single photon counting mode by PC-based TCSPC electronics (DPC 230, Becker & Hickl, Berlin). The focus of the trapping laser, which was operated at 785 nm (Sanyo laser diode; Thorlabs), was overlaid with the confocal laser by beam steering mirrors. To determine trapping, we analyzed the photon burst length (or diffusion times of single beads through the confocal detection volume, respectively). Because the repulsive part of the optical tweezers forces points along the propagation axis of the trapping laser, we mechanically restricted the movement of the particles in z direction using a flat trapping chamber with a maximum height of less than 2 μm.

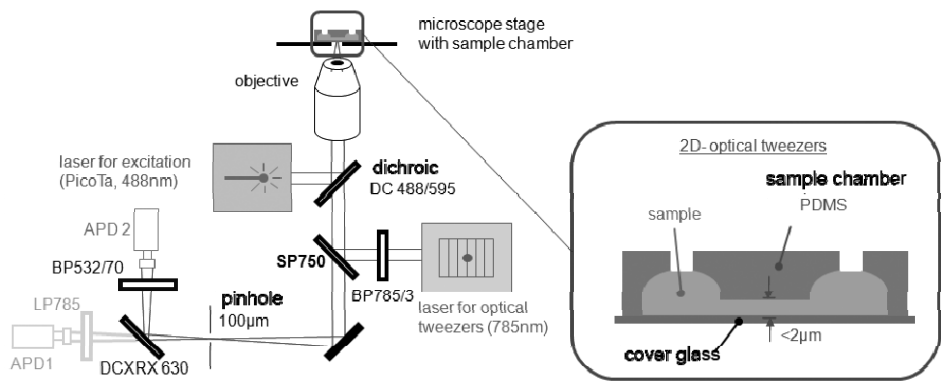

**Figure 1. Combination of optical tweezers and a confocal microscope setup**. The 785 nm laser diode is used for the optical tweezers, and a second laser at 488 nm is exciting the fluorescent particle simultaneously in the overlapping foci. The flat sample chamber with a maximum height of 2 μm limits diffusion of the particles to two dimensions. APD1 is used for alignment of the two lasers, fluorescene is detected by APD2.

The slowest diffusion times for the 100 nm-sized polystyrene beads increased from 71 ms to 404 ms (Figure 2) at a low trapping power of 21 mW (785 nm).

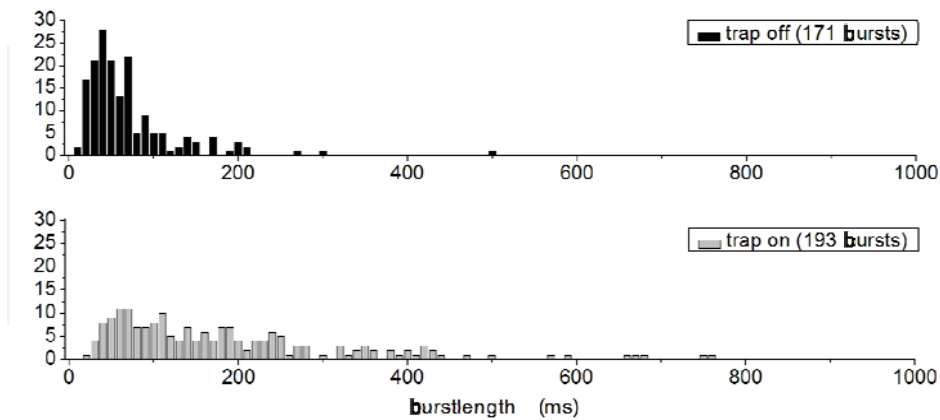

**Figure 2. Diffusion times of 100 nm polystyrene beads** without (upper histogram) and with (lower histogram) optical tweezers. At a trapping power of 21 mW the maximum burst lengths increased by a factor 5.7 .

## 3.2 Monte Carlo simulations of the optical trap

We simulated the diffusion of a 100 nm bead in Cartesian three dimensional space. In x and y direction we chose periodic boundaries; in z direction we limited the movement of the bead. The intensity profiles of the optical tweezers, the confocal excitation and the detection volume were Gaussians, and their maximum was located in the origin of the coordinate system. To simulate the movement of the bead from position $x_i$ to the next position $x_{i+1}$, we rewrite eq (3):

$$x_{i+1}(\Delta t) = \frac{1}{\gamma} \cdot F_{opt}^{(0)}(x_i) \cdot \Delta t + \Delta x_s^1 + x_i \qquad (11)$$

We used the distribution obtained in (4) as $p(x_s^i, \Delta t)$ to get a randomized value for $x_s^i$ for in the $i$-th time step, an integer multiple of the step size $\Delta t$ of the simulation. $F_{opt}^{(0)}$ is the zero order coefficient of the Taylor expansion of the optical tweezers force. To use (11) we have to verify that the condition $\Delta t / \tau \gg 1$ is satisfied: for a spherical 100 nm sized particle with a mass of m = $7.3 \cdot 10^{-22}$ kg and a damping constant of $\gamma$ = 9.4 $10^{-24}$ kg/s in water we get $\Delta t / \tau$ = $2.57 \cdot 10^5 \gg 1$ for $\Delta t = 2 \cdot 10^{-7}$ s. We also verify, that the zero order term of the optical tweezers forces describes the field strength good enough: given a typical field strength of 1 pN and a maximum force gradient of 1 pN/µm, the contributions of the zero order term is 10 times higher than the first order.

We chose the beam waist $\omega_0$ of the trapping laser (at 785 nm) to be 392.5 nm with a Rayleigh length of 1.57 µm for the laser focus (intensity drop to $1/e^2$). For the excitation laser we chose a beam waist of 244 nm and a Rayleigh length of 967 nm. The bead itself was 100 nm in size with a refractive index of 1.57 (polystyrene) in water (refractive index 1.33). The simulation volume was a box with 6 µm x 6 µm width (x, y dimensions) and 2 µm high (-1 µm to 1 µm in z). The simulation temperature was 300 K.

Figure 3 shows the result of the simulation. All fluorescence traces through the confocal detection volume had been averaged to a mean burst length for one specific trapping laser power. The optical tweezers increased the mean burst length above a threshold of 8 mW. Accordingly, the attractive potential of the trap is $-6.22 \cdot 10^{-22}$ J, and the thermal energy of the particle at 300 K is $6.22 \cdot 10^{-22}$ J. A minimum force has to be applied to overcome the thermal energy of the particle.

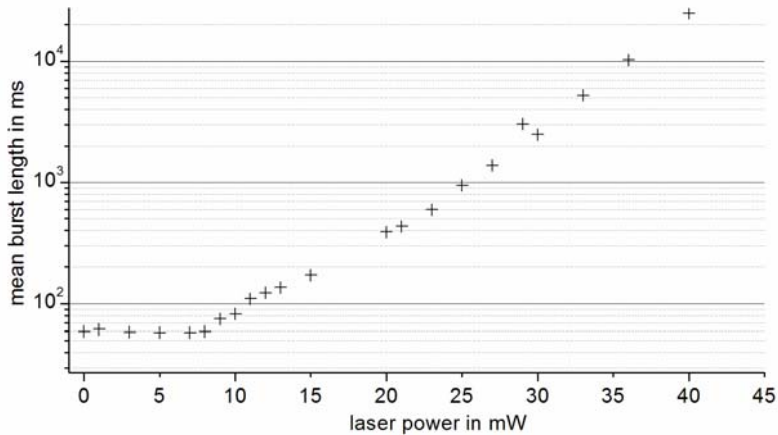

**Figure 3. Simulations of the photon burst lengths** (transit times through the confocal detection volume) of fluorescent 100 nm polystyrene beads hold by two dimensional optical tweezers at different laser powers.

The average refractive index of a proteoliposome (with $F_oF_1$-ATP synthase reconstituted in a liposome) was estimated to 1.37. Therefore, a 5.8 higher force is needed to get the same attractive potential as for polystyrene beads. Because of the larger liposome diameter of 131 nm, we need 45% less laser power than for 100 nm vesicles. In summary at least 21 mW (8 mW x 5.8 x 0.45) trapping laser power is needed to increase the diffusion time of proteoliposomes. To increase the diffusion time to one second about 65 mW are needed. This laser power is too high for simultaneous FRET measurements on the enzyme because the trap laser will instantaneously bleach the FRET labels. It also effects on the functionality of the protein, and, therefore, optical tweezers are not suitable for simultaneous measuring the conformational changes and rotary motions in $F_oF_1$-ATP synthase *via* FRET.

# 4 THE ABELTRAP

## 4.1 PDMS sample chamber preparation

The ABELtrap is a microfluidic system with a cross structure. The center of the trapping region should be as thin as possible, i.e. less than 1 µm, to constrain Brownian motion. Each of the 4 channels of the cross structure contains one platinum electrode. The 4 channels are connected by a ring structure with an average height of 40 µm. We produced different chamber structures to test the trapping precision and simulated the electric fields within these chambers.

The microfluidic chamber was made from polydimethylsiloxane (PDMS) bound on a glass cover slide. To fabricate the chamber systems we used standard photolithography. Figure 4 shows the process of production of the PDMS part: (1 and 2a) we spin-coat a photo resist mixture (50% SU-8, type 2035, MicroChem; 50% methoxy propyl acetate) onto the silicon waver at 6000 rpm and expose the cross structure with UV light (3a). (4a) After developing the resist (Dev-600, MicroChem) we spin-coat a second layer (1b, 2b) at 1000 rpm (SU-8, type 2035, MicroChem) onto the waver to produce the ring structure. After UV exposure and development (4b) of the ring structure, we have created the chamber stamp. Finally we use PDMS (Sylgard® 184 Elastomer Kit; Dow Corning) with a hardener to make the micro fluidic chamber systems in a petri dish (5). The cross structure or trapping region has a height of 600 nm, and the ring structure an average height about 40 µm.

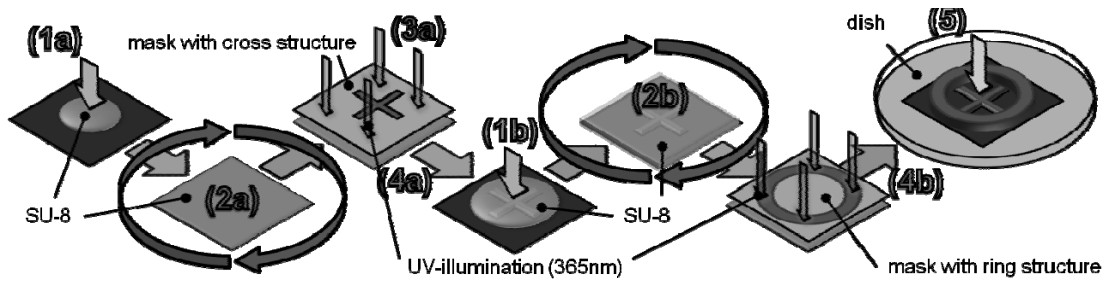

**Figure 4. Fabricating the ABELtrap chamber with PDMS (see text).**

## 4.2 Different trapping regions and chamber designs

We tested three different chamber designs. The first chamber had a cross structure with very long channels (323 µm) to increase the fabrication tolerances. For the second chamber we reduced the channel lengths to 123 µm to increase the field strength in the trapping region. The third chamber was a copy of the original designs by Adam E Cohen, and the channels of the cross structure had a length of only 31.5 µm. For the production of this chamber a special alignment system was needed because the tolerances for overlapping the cross- and the ring structures are below 6 µm.

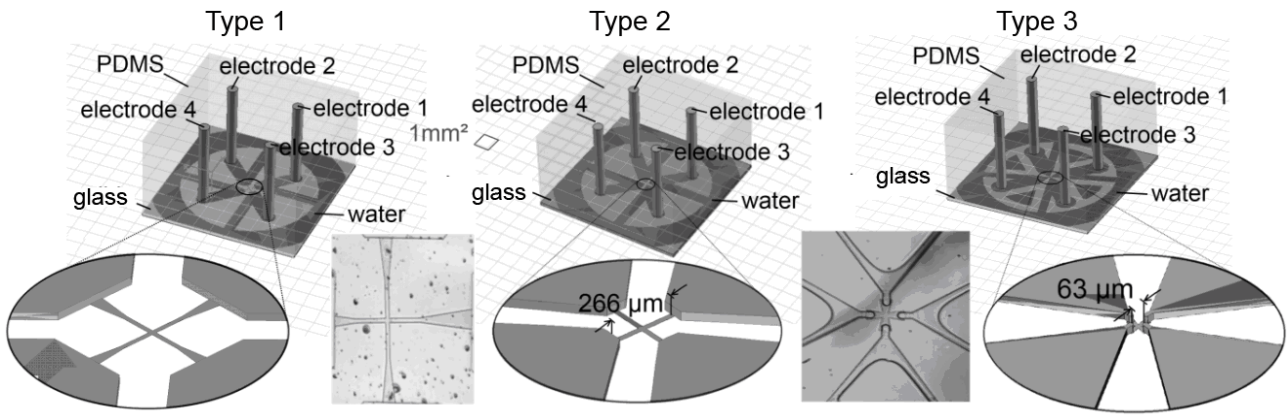

**Figure 5. Three ABELtrap chamber designs**. Chamber versions type 1 (left), type 2 (middle), and type 3 (right). Bright field images with the different channel lengths are shown for types 1 and 3.

### 4.3 Simulations of electrical fields

To find the optimal chamber design, we simulated the electrical fields within the different chamber designs. The maximum feedback cycle time we used was 700 Hz. Therefore we chose a direct current (dc) simulation (software: CST studio) to model the electrical fields. Hydration was not considered. A probe line crossing the trapping region was used for visualization of the electric fields in Figure 6. The trapping regions of all chambers were comparable (version 1 and 2: 20μm x 20 μm, version 3: 18 μm x 18μm). Figure 6 shows the absolute values and gradients of the electric fields in the trapping regions when applying an electric potential of +2 V on electrode 1, and -2 V on electrode 3. Electrodes 2 and 4 had been grounded.

Running the ABELtrap, electrodes 1 and 3 will act as a coupled pair to move the vesicle along the axis between electrodes 1 and 3. Accordingly electrodes 2 and 4 act as a pair, but both pairs operate independently in the symmetric chambers. For chamber 1, the average intensity is 618 μV/μm; for version 2 it is 816 μV/μm, and for type 3 it is 1119 μV/μm. The electrical fields vary from the borders of the trapping region to the center for chamber type 1 about 2%, for type 2 about 2.8%, and for type 3 about 9.3%.

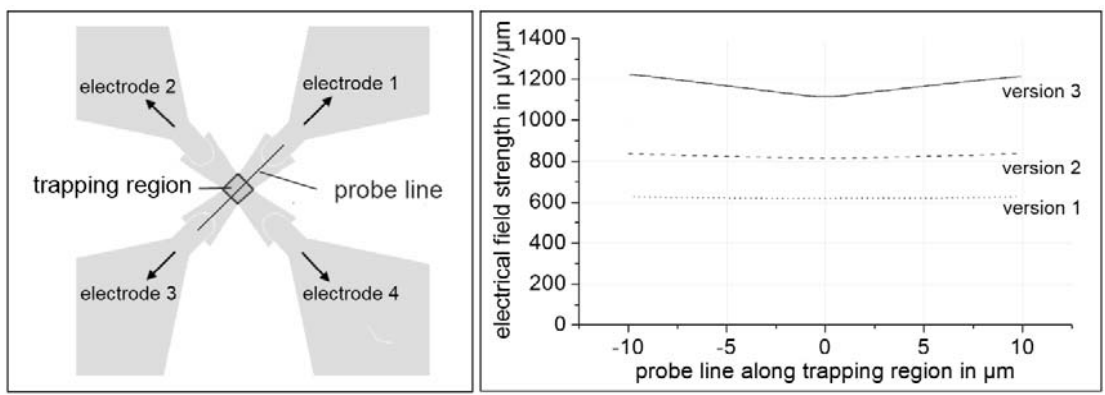

**Figure 6. Electrical field simulations.** Right, position of the probe line through the trapping region. Left, absolute electrical fields around the center of the trapping region.

### 4.4 ABELtrap, microscope setup and correction algorithm

Figure 7 shows the combination of the ABELtrap (gray parts, EMCCD-based, the so-called "software"-ABELtrap) with a confocal microscope setup (white parts, only one APD is shown here).

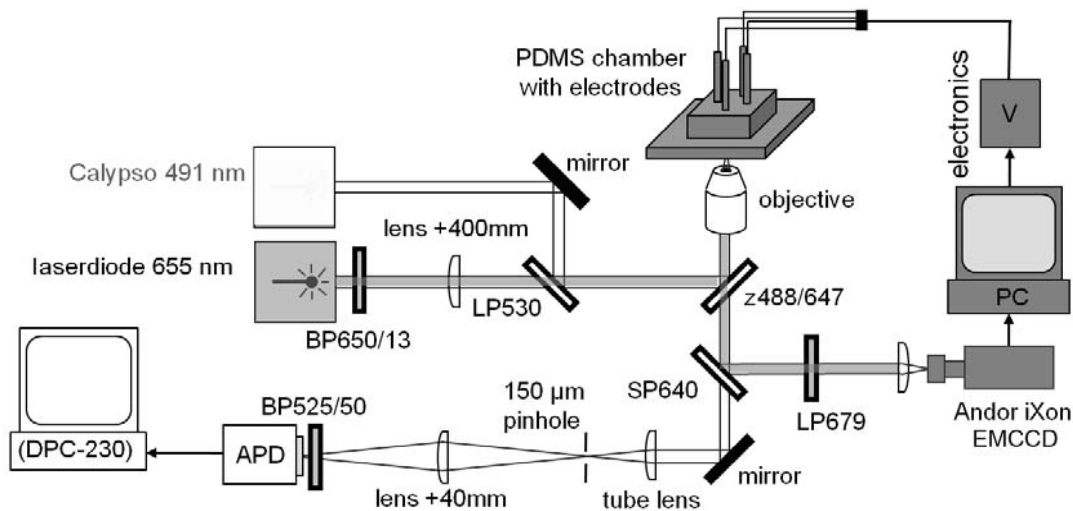

**Figure 7. Combination of an ABELtrap (gray parts) with a confocal microscope setup (white parts).** The system is built on an Olympus IX 71 stage using a 100x TIRF oil objective with NA 1.49 (see text for details).

As the test sample, we used dual-labeled fluorescent liposomes containing about 10 ATTO488-DOPE lipids and about 60 ATTO680-DOPE lipids (ATTO-TEC). The red label (i.e. ATTO680) was excited by the laser diode in wide-field illumination mode around the center of the trapping region. A fast EMCCD camera (iXon$^+$ 897, Andor) detected the fluorescence of Atto680. The computer calculated the voltages to correct the particle position with respect to the center of the confocal volume of the second, confocal, blue laser at 491 nm (Calypso, Cobolt; von Gegerfelt Photonics). A D/A converter (NI-6713, National Instruments) and custom-made amplifier electronics were used to apply the potentials to the platinum electrodes.

The ABELtrap control software was written in Labview (National Instruments). The actual frame of the EMCCD camera was grabbed and the background was subtracted. The camera was running as fast as possible in the "internal triggering" mode. The corrected images were analyzed with a simple intensity threshold algorithm: all pixels above a given intensity threshold were set to "active". All active pixel that were connected formed an area. Areas smaller than a minimum area size were ignored. Thus we were able to run the ABELtrap even at a low signal-to-noise ratio, that is, when the pixel noise had a similar magnitude compared to the intensity of the pixels of the fluorescent liposomes. Then the centroid of all active areas was calculated. The algorithm was fast enough to calculate the positions within a fraction of the used cycle time. The vesicle localized next to the confocal volume was trapped. We transformed the position into a voltage using equation (8) and assumed that the electrical field depended linearly on the potential in x ($U_x$) and y ($U_y$) direction, i.e. using eqs. (12):

$$U_x = \frac{k}{\tau}\Delta x \quad and \quad U_y = \frac{k}{\tau}\Delta y \tag{12}$$

with $k$, constant prefactor that can be adjusted by minimizing the mean square displacement of the trapped particle, $\tau$, the frame rate of the camera, and $\Delta x$ and $\Delta y$, displacements of the vesicle to the set point. We saved the raw data to calculate and control the particle displacements after the ABELtrap experiments.

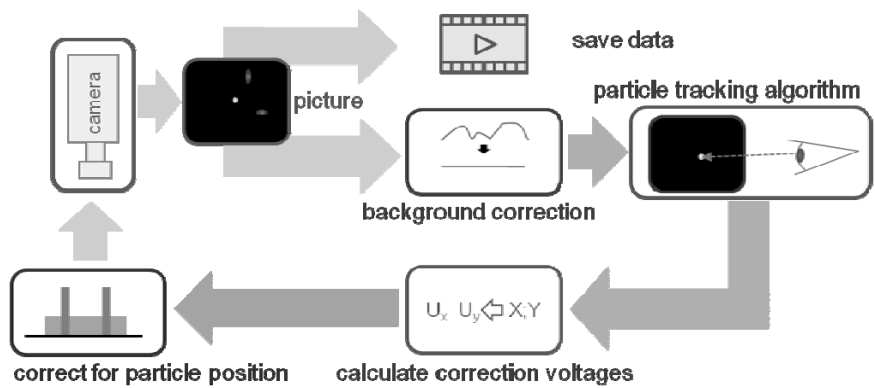

**Figure 8. Flow chart of the correction algorithm of the ABELtrap.**

**4.5 Analyzing particle displacements in the ABELtrap for different camber designs**

To optimize the chamber geometries we recorded the x and y positions of fluorescent particles in the different chamber types and computed position deviation histograms. The variances of the distributions were determined by Gaussian fittings and compared. All factors causing increased variances were minimized or held constant during the measurements. We used fluorescent beads (100 nm) with high signal-to-noise ratio (SNR) to measure the localization. Figure 9 shows a simulation of the localization precision depending on the camera frame rates and the SNR. For the simulations we assumed Gaussian distributed fluorescence intensity profiles and a pixel size of 380 nm. The simulations also included the "blooming effect" of a moving object due to a detection system with finite exposure times. As a result a lower limit was found for the precision of the localization which was due to the finite pixel size of the CCD. However, 200 nm precision for trapping seemed to be achievable even at low frame rates of 100 Hz and with a given low SNR of 2.

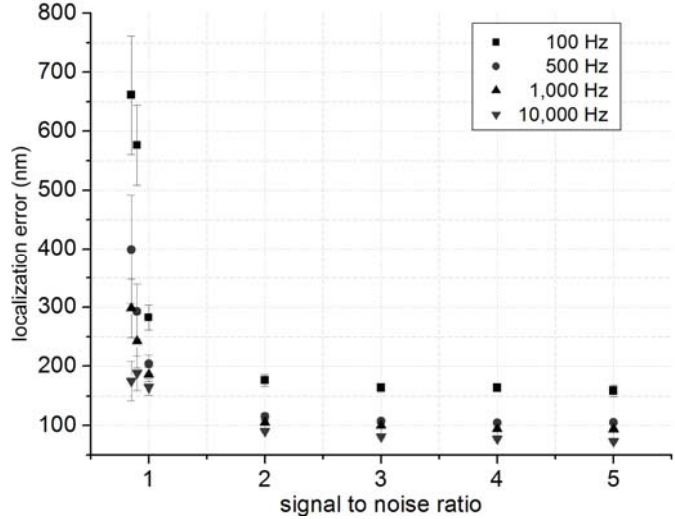

**Figure 9. Simulation of the localization error** depending on the signal-to-noise ratio (SNR) of the "fluorescent" particle. The localization error converts to a finite minimum value for high SNR.

Following the simulations we made several localization measurements of fluorescent beads in the ABELtrap with different chosen prefactors *k* (see equation (12)). Those with the lowest variances for chamber types 1 and 3 are summarized in Table 1. To optimize trapping performance we also varied the pixel binning. This is an internal hardware mode of the EMCCD camera which decreases the readout time for one frame. Because the resulting "super pixel" is the sum of more pixels, binning increased also the SNR. However, binning also decreased the localization precision.

|  | Chamber type 1 |  | Chamber type 3 |  |
|---|---|---|---|---|
| frame rate | 400 Hz | 238 Hz | 400Hz | 236Hz |
| pixel-binning: | 4x4 | 2x2 | 4x4 | 2x2 |
| $\sigma_{all}$ | 400 nm | 410 nm | 210 nm | 240 nm |
| $d_t$ | 148 nm | 192 nm | 148 nm | 192 nm |
| $\sigma_\Delta$ | 372 nm | 362 nm | 149 nm | 144 nm |

**Table 1.** Standard deviations of 100 nm beads in the ABELtrap for different chamber designs.

Comparing $\sigma_\Delta$ for the different chamber types we found that the standard deviation $\sigma_\Delta$ for type 1 (with the long channels towards the trapping region) is more than two times larger than for type 3. Finally we optimized the feedback control of the ABEL-trap, i.e. the feedback loop time and the timing for the feedback. Both have to be constant from frame to frame of the EMCCD. It turned out, that a good trigger source for the feedback control is the "fire" output of the EMCCD camera (the camera status output using TTL).

**4.6 Analyzing the ABELtrap performance for a doubly-labeled liposome**

Our ultimate goal is to perform extended confocal FRET measurements on a single $F_oF_1$-ATP synthase that had been reconstituted in a labeled liposome. As the test sample we used ATTO680-DOPE containing liposomes to evaluate the improvements of the ABELtrap precision. Figure 10 shows the result of the optimization for trapping ATTO680-DOPE stained liposomes with an estimated average of sixty ATTO680-DOPE (plus ten ATT0488-DOPE lipids) per liposome with a mean diameter of 130 nm. In Figure 10a, the two dimensional histogram of the relative localization of the particle is shown. All x and y displacements were summed up in a one dimensional histogram (Figure 10 b). Fitting with a Gaussian resulted in standard deviation $\sigma_{all}$ of 225 nm.

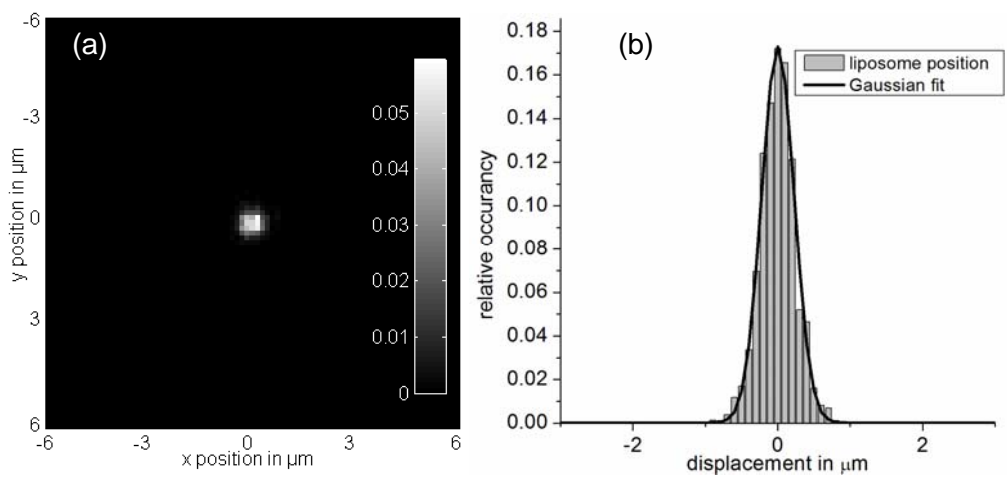

**Figure 10**. **Position analysis of a trapped double labeled liposome**. (**a**), two dimensional histogram of the relative particle position. (**b**), displacement distribution in a one dimensional histogram summing all displacements in x and y direction**.**

The confocal detection volume had an estimated diameter of 1 μm ($1/e^2$ threshold) corresponding to an acceptable maximum limit for the standard deviation of 250 nm in the ABELtrap. Experimentally the liposome had been trapped over 3360 frames with a feedback loop of 287 Hz. This liposome was trapped for more than 10 seconds.

Finally we probed the overlap of the confocal detection volume with the set point of the ABELtrap. The second fluorescent lipid ATTO488-DOPE in the liposome is excited with 491 nm by the confocal laser. ATTO680 fluorescence is used to localize the liposome. When a liposome entered the confocal detection volume, fluorescence of ATTO488 is detected by an avalanche photodiode. Stepwise photobleaching of ATTO488-DOPE was observed (Figure 11).

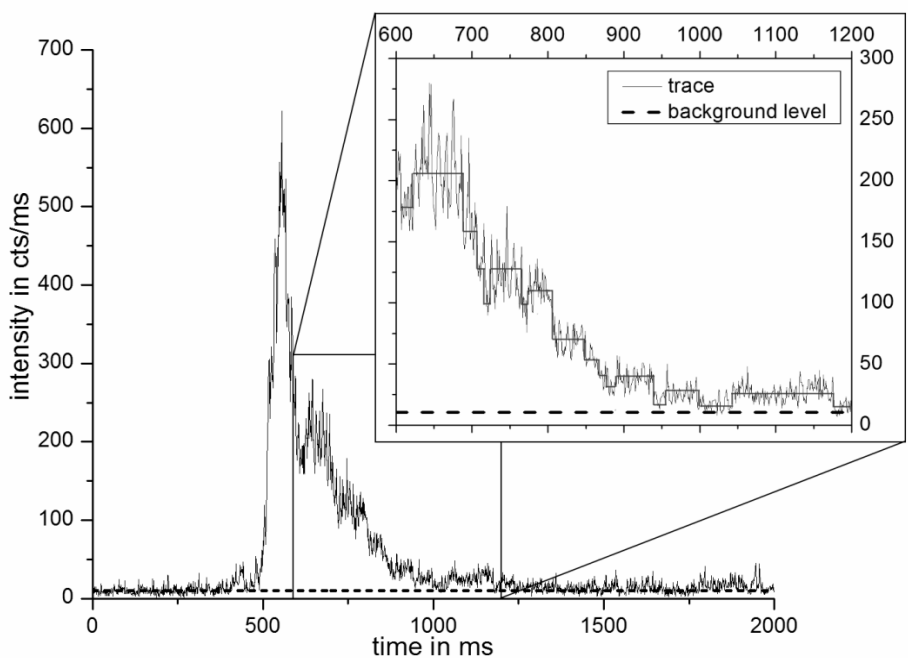

**Figure 11**. **Stepwise photobleaching of ATTO488-DOPE in an ATTO680-DOPE stained liposome hold by the ABELtrap.** The excitation intensity of the confocal laser at 491 nm was 50 μW. The inset shows bleaching and blinking of few ATT0488-DOPE lipids in the liposome over an observation time of 800 ms.

The inset in Figure 11 clearly shows the stepwise ATTO488 photobleaching over 800 ms and additional blinking behavior of the ATTO488 fluorophores. Here, steps were assigned manually to guide the reader.

## 5 OUTLOOK

The ABELtrap using a EMCCD to localize a fluorescent liposome is a promising manipulation tool for single-molecule studies of membrane proteins. The laser powers needed to excite the fluorescently labeled liposome and the embedded membrane protein is several orders of magnitude smaller than those in optical traps. This allows to observe additional single fluorophores which are required for FRET-based studies of the conformational changes of a membrane protein. Our aim to increase the observation time of liposome-reconstituted $F_oF_1$-ATP synthases in a confocal FRET experiment seems to be achievable in the ABELtrap as evaluated here.

The newer versions of the ABELtrap developed in the laboratories of W. E. Moerner (Stanford) and A. E. Cohen (Harvard) operate at much higher feedback rates, that is, in the 20 kHz range. Accordingly, single soluble proteins with sizes in the 10 nm range and even single fluorophores can be trapped for more than a second. The intensity fluctuations of a single fluorescent lipid in a liposome as seen in Figure 11 will be dramatically reduced. We expect to resolve details of the conformational dynamics of $F_oF_1$-ATP synthase and interactions of different inhibitors with subunit rotation in the near future.

**Acknowledgement**
M.B. wants to thank A. E. Cohen and W. E. Moerner for their ongoing generous support for measuring the rotary motions in $F_oF_1$-ATP synthase in the ABELtrap. Their ABELtrap concept is inspiring since the first presentation. Financial support by the Baden-Württemberg Stiftung (project P-LS-Meth/6 in the program "Methods for Life Sciences") is acknowledged.

## 6 References

[1] H. Noji, R. Yasuda, M. Yoshida, K. Kinosita, Jr., *Nature* **1997**, 386, 299.
[2] M. Yoshida, E. Muneyuki, T. Hisabori, *Nat Rev Mol Cell Biol* **2001**, 2, 669.
[3] P.D. Boyer, *Annu Rev Biochem* **1997**, 66, 717.
[4] M. Borsch, M. Diez, B. Zimmermann, R. Reuter, P. Graber, *FEBS Lett* **2002**, 527, 147.
[5] M. Borsch, M. Diez, B. Zimmermann, M. Trost, S. Steigmiller, P. Graber, *Proc. SPIE* **2003**, 4962, 11.
[6] M. Diez, B. Zimmermann, M. Borsch, M. Konig, E. Schweinberger, S. Steigmiller, R. Reuter, S. Felekyan, V. Kudryavtsev, C.A. Seidel, P. Graber, *Nat Struct Mol Biol* **2004**, 11, 135.
[7] B. Zimmermann, M. Diez, N. Zarrabi, P. Graber, M. Borsch, *Embo J* **2005**, 24, 2053.
[8] N. Zarrabi, B. Zimmermann, M. Diez, P. Graber, J. Wrachtrup, M. Borsch, *Proc. SPIE* **2005**, 5699, 175.
[9] B. Zimmermann, M. Diez, M. Borsch, P. Graber, *Biochim Biophys Acta* **2006**, 1757, 311.
[10] N. Zarrabi, M.G. Duser, S. Ernst, R. Reuter, G.D. Glick, S.D. Dunn, J. Wrachtrup, M. Borsch, *Proc. SPIE* **2007**, 6771, 67710F.
[11] N. Zarrabi, M.G. Duser, R. Reuter, S.D. Dunn, J. Wrachtrup, M. Borsch, *Proc. SPIE* **2007**, 6444, 64440E.
[12] M.G. Duser, Y. Bi, N. Zarrabi, S.D. Dunn, M. Borsch, *J Biol Chem* **2008**, 283, 33602.
[13] M.G. Duser, N. Zarrabi, D.J. Cipriano, S. Ernst, G.D. Glick, S.D. Dunn, M. Borsch, *Embo J* **2009**, 28, 2689.
[14] K.M. Johnson, L. Swenson, A.W. Opipari, Jr., R. Reuter, N. Zarrabi, C.A. Fierke, M. Borsch, G.D. Glick, *Biopolymers* **2009**, 91, 830.
[15] S. Steigmiller, B. Zimmermann, M. Diez, M. Borsch, P. Graber, *Bioelectrochemistry* **2004**, 63, 79.
[16] T. Krebstakies, B. Zimmermann, P. Graber, K. Altendorf, M. Borsch, J.C. Greie, *J Biol Chem* **2005**, 280, 33338.
[17] N. Zarrabi, S. Ernst, M.G. Dueser, A. Golovina-Leiker, W. Becker, R. Erdmann, S.D. Dunn, M. Borsch, *Proc. SPIE* **2009**, 7185, 718505.
[18] M. Borsch, P. Graber, *Biochemical Society Transactions* **2005**, 33, 878.
[19] M. Borsch, *Biological Chemistry* **2011**, 392, 135.
[20] S. Fischer, P. Graber, P. Turina, *J Biol Chem* **2000**, 275, 30157.
[21] S. Steigmiller, P. Turina, P. Graber, *Proc Natl Acad Sci U S A* **2008**, 105, 3745.
[22] T. Heitkamp, R. Kalinowski, B. Bottcher, M. Borsch, K. Altendorf, J.C. Greie, *Biochemistry* **2008**, 47, 3564.
[23] N. Zarrabi, T. Heitkamp, J.-C. Greie, M. Borsch, *Proc. SPIE* **2008**, 6862, 68620M.
[24] S. Ernst, C. Batisse, N. Zarrabi, B. Bottcher, M. Borsch, *Proc. SPIE* **2010**, 7569, 75690W.
[25] A. Armbruster, C. Hohn, A. Hermesdorf, K. Schumacher, M. Borsch, G. Gruber, *FEBS Lett* **2005**, 579, 1961.

[26]	M. Diepholz, M. Borsch, B. Bottcher, *Biochem Soc Trans* **2008**, 36, 1027.
[27]	I.B. Schafer, S.M. Bailer, M.G. Duser, M. Borsch, R.A. Bernal, D. Stock, G. Gruber, *J Mol Biol* **2006**, 358, 725.
[28]	F.M. Boldt, J. Heinze, M. Diez, J. Petersen, M. Borsch, *Anal Chem* **2004**, 76, 3473.
[29]	F.E. Alemdaroglu, S.C. Alexander, D.M. Ji, D.K. Prusty, M. Borsch, A. Herrmann, *Macromolecules* **2009**, 42, 6529.
[30]	K. Ding, F.E. Alemdaroglu, M. Borsch, R. Berger, A. Herrmann, *Angew Chem Int Ed Engl* **2007**, 46, 1172.
[31]	F.E. Alemdaroglu, J. Wang, M. Borsch, R. Berger, A. Herrmann, *Angew Chem Int Ed Engl* **2008**, 47, 974.
[32]	C. Winnewisser, J. Schneider, M. Borsch, H.W. Rotter, *Journal of Applied Physics* **2001**, 89, 3091.
[33]	M. Borsch, P. Turina, C. Eggeling, J.R. Fries, C.A. Seidel, A. Labahn, P. Graber, *FEBS Lett* **1998**, 437, 251.
[34]	J. Ufheil, F.M. Boldt, M. Borsch, K. Borgwarth, J. Heinze, *Bioelectrochemistry* **2000**, 52, 103.
[35]	S. Steigmiller, M. Borsch, P. Graber, M. Huber, *Biochim Biophys Acta* **2005**, 1708, 143.
[36]	M. Borsch, *Proc. SPIE* **2010**, 7551, 75510G.
[37]	M. Borsch, R. Reuter, G. Balasubramanian, R. Erdmann, F. Jelezko, J. Wrachtrup, *Proc. SPIE* **2009**, 7183, 71832N.
[38]	M. Diez, M. Borsch, B. Zimmermann, P. Turina, S.D. Dunn, P. Graber, *Biochemistry* **2004**, 43, 1054.
[39]	E. Galvez, M. Duser, M. Borsch, J. Wrachtrup, P. Graber, *Biochem Soc Trans* **2008**, 36, 1017.
[40]	F. Neugart, A. Zappe, D.M. Buk, I. Ziegler, S. Steinert, M. Schumacher, E. Schopf, R. Bessey, K. Wurster, C. Tietz, M. Borsch, J. Wrachtrup, L. Graeve, *Biochim Biophys Acta* **2009**, 1788, 1890.
[41]	G. Modesti, B. Zimmermann, M. Borsch, A. Herrmann, K. Saalwachter, *Macromolecules* **2009**, 42, 4681.
[42]	J. Tisler, G. Balasubramanian, B. Naydenov, R. Kolesov, B. Grotz, R. Reuter, J.P. Boudou, P.A. Curmi, M. Sennour, A. Thorel, M. Borsch, K. Aulenbacher, R. Erdmann, P.R. Hemmer, F. Jelezko, J. Wrachtrup, *Acs Nano* **2009**, 3, 1959.
[43]	K. Peneva, G. Mihov, A. Herrmann, N. Zarrabi, M. Borsch, T.M. Duncan, K. Mullen, *J Am Chem Soc* **2008**, 130, 5398.
[44]	A.E. Cohen, W.E. Moerner, *Proc. SPIE* **2005**, 5699, 296.
[45]	A.E. Cohen, W.E. Moerner, *Proc. SPIE* **2005**, 5930, 59300S.
[46]	A.E. Cohen, W.E. Moerner, *Proc Natl Acad Sci U S A* **2006**, 103, 4362.
[47]	Y. Jiang, Q. Wang, A.E. Cohen, N. Douglas, J. Frydman, W.E. Moerner, *Proc. SPIE* **2008**, 7038, 703807.
[48]	A.E. Cohen, W.E. Moerner, *Opt Express* **2008**, 16, 6941.
[49]	P.A.M. Neto, H.M. Nussenzveig, *EPL (Europhysics Letters)* **2000**, 50, 702.
[50]	A. Rohrbach, E.H. Stelzer, *J Opt Soc Am A Opt Image Sci Vis* **2001**, 18, 839.
[51]	K. Svoboda, S.M. Block, *Annu Rev Biophys Biomol Struct* **1994**, 23, 247.
[52]	A.E. Cohen, *Phys Rev Lett* **2005**, 94, 118102.